\documentstyle[aps,prl,epsf,multicol]{revtex}

\newcommand{\bleq}{\ifpreprintsty
                   \else
                   \end{multicols}\vspace*{-3.5ex}{\tiny 
                   \noindent\begin{tabular}[t]{c|}
                   \parbox{0.493\hsize}{~} \\ \hline \end{tabular}}
                   \fi}
\newcommand{\eleq}{\ifpreprintsty
                   \else
                   {\tiny\hspace*{\fill}\begin{tabular}[t]{|c}\hline
                    \parbox{0.49\hsize}{~} \\ 
                    \end{tabular}}\vspace*{-2.5ex}\begin{multicols}{2}
                    \fi}
\newcommand{\bcols}{\ifpreprintsty\else\begin{multicols}{2}\fi}
\newcommand{\ecols}{\ifpreprintsty\else\end{multicols}\fi}

\begin{document}
\draft
\title{Effects of electron-electron interaction on the conductance
 of open quantum dots}

\author{P. W. Brouwer$^a$ and I. L. Aleiner$^b$}
\address{$^a$ Lyman Laboratory of Physics, Harvard University, 
  Cambridge, MA 02138\\
  $^b$ Department of Physics and Astronomy, SUNY at Stony Brook,
  Stony Brook, NY 11794\\
{\rm (\today)}
\medskip ~~\\ \parbox{14cm}{\rm
We study the effect of electron-electron interaction on the conductance
of open quantum dots. We find that Coulomb interactions (i) do not
affect the ensemble averaged conductance $\langle G \rangle$ if
time-reversal symmetry has been broken by a magnetic field, (ii)
enhance weak localization and weak anti-localization corrections to
$\langle G \rangle$ in the absence of a magnetic field, (iii) increase
conductance fluctuations, and (iv) enhance the effect of short
trajectories on the conductivity of quantum dot.
}}
\maketitle \bigskip


\bcols

The phenomenon of Coulomb blockade in quantum dots is commonly
associated with dots that are coupled to the outside world via
tunneling contacts \cite{AL}. In these systems, the total charge on the dot is
quantized in units of the electron charge $e$. Once the temperature $T$
becomes smaller than the charging energy $E_C$ (the energy needed to
add an extra electron to the system), transport through the quantum dot
is suppressed, unless the system is tuned to a charge degeneracy
point. On the other hand, if the connection to the outside
world is via ballistic point contacts with a conductance much larger
than the conductance quantum $2 e^2/h$, charge is no longer quantized,
and the Coulomb blockade is lifted.

In recent years, it has become possible to study the intermediate
regime, of quantum dots with ballistic point contacts with only a few
propagating channels at the Fermi level, so that their conductance is
$\sim e^2/h$. For temperatures $T$ comparable to the spacing $\Delta$
of single-particle levels in the quantum dot, the conductance of these
open quantum dots shows mesoscopic fluctuations of the same order as
the average \cite{Marcus,Westervelt}.  Theoretically, the large
mesoscopic fluctuations are understood within the framework of
random-matrix theory \cite{BarangerMello,PEI,Efetov,BReview}, which
employs a picture of non-interacting electrons, and does not account
for Coulomb blockade effects.  A partial justification for this
approach can be obtained from the work of Furusaki and Matveev
\cite{FM}, who calculated the effect of electron-electron interactions
on the conductance $G$ of a quantum dot with ballistic single-channel
point contacts in the limit $E_C \gg T \gg \Delta$, where there are no
mesoscopic fluctuations. They found $G=e^2/h$, indicating that there is
no Coulomb blockade effect on the conductance for ballistic point
contacts in this limit. At the same time, the presence of weakly
reflecting tunnel barrier in the contact will drive the system into a
state of Coulomb blockade, in which transport is inhibited.

That this is not the complete picture was explained in a recent work by
one of the authors and Glazman \cite{AG}, who showed that as a result
of the interplay of mesoscopic fluctuations and electron-electron
interactions, Coulomb blockade effects on the capacitance and the
tunneling density of states of a quantum dot persist, even at a
perfect transparency of the point contacts. 
This so-called ``mesoscopic charge quantization'' urges us to
reconsider the effect of Coulomb interactions on the conductance
of a quantum dot with ideal point contacts. 
In this letter we report a calculation the conductance of such a
quantum dot and its mesoscopic
fluctuations, in the presence of electron-electron interactions.

\begin{figure}
\epsfxsize=0.6\hsize
\hspace{0.15\hsize}
\epsffile{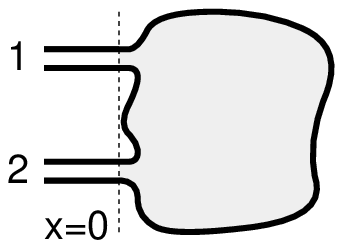}
\refstepcounter{figure}
\label{fig:1}

{FIG.\ \ref{fig:1}. Quantum dot (dotted) connected to two ideal
leads (labeled $1$ and $2$).}
\end{figure}

We consider a quantum dot connected to electron reservoirs
via two leads (labeled $1$ and $2$), see Fig.\ \ref{fig:1}. In each of
the leads there are $N$ propagating channels at the Fermi level $E_F$.
For the Hamiltonian of the system we take
\begin{mathletters} \label{eq:model}
\begin{equation}
  {\cal H} = {\cal H}_{F} + {\cal H}_{C},
\end{equation}
where ${\cal H}_{F}$ is the Hamiltonian of the non-interacting electrons
and ${\cal H}_{C}$ is the interaction Hamiltonian, corresponding to a
capacitive interaction inside the quantum dot,
\begin{eqnarray}
  {\cal H}_{F} &=& \sum_{\sigma = \pm} \int d \vec r\, \psi^{\dagger}_{\sigma} \left( -{1 \over 2 m} \nabla^2 + U(\vec r) - \mu \right) \psi_{\sigma}^{\vphantom{Please remove this superscript!}} , \\
  {\cal H}_{C} &=& {E_C \over 2 e^2} \left( Q - n e \right)^2,\ \
    Q = e \sum_{\sigma = \pm} \int_{x > 0} d\vec r\, \psi^{\dagger}_{\sigma} \psi_{\sigma}^{\vphantom{Please remove this superscript!}}.
  \label{eq:HamC}
\end{eqnarray}
\end{mathletters}%
\noindent
The index $\sigma$ denotes spin. The potential $U(\vec r)$ describes
the confinement by the gates surrounding the quantum dot and the
leads.  The leads are located at $x < 0$, and the dot at $x > 0$, see
Fig.\ \ref{fig:1}. As is explained in Ref.\ \onlinecite{AG}, the
interaction Hamiltonian ${\cal H}_{C}$ with the capacitive interaction
provides a sufficient description of the electron-electron interactions
in a metal of semiconductor quantum dot if the electrons explore the
dot ergodically before they exit the dot through one of the point
contacts, i.e.\ if all the involved energy scales are much smaller than
the Thouless energy of the dot $E_T$.  It is in this regime that the
conductance statistics become universal and random-matrix theory can be
used as a model for the non-interacting Hamiltonian ${\cal H}_F$
\cite{BReview}.

To summarize the result of our calculation, which is described below,
we find that the presence of the interaction Hamiltonian
(\ref{eq:HamC}) enhances mesoscopic (quantum-interference)
corrections to the conductance.  In particular:

(i).
There is no interaction correction for the ensemble averaged conductance
$\langle G \rangle$ in the presence of a time-reversal symmetry
breaking field (unitary ensemble). This is different from disordered bulk
systems, where interactions do have an effect on the conductance, even in
the presence of a weak magnetic field \cite{BulkInteractions}.

(ii).
In the absence of a magnetic field (orthogonal ensemble), the weak-localization correction to $\langle G \rangle$ is enhanced by the interactions. We have computed $\langle G \rangle$ in a perturbation series in $N \Delta/T$, where $N$ is the number of channels in each of the point contacts and $\Delta$ is the mean level spacing in the quantum dot,
\begin{eqnarray} \label{eq:WL}
  \langle G \rangle &=&
     {N \over 2} - {N \over 4 N + 2} -
     {c_{N} \over N}{N \Delta \over 8 \pi^2 T}.
\end{eqnarray}
Here $c_N$ is a numerical constant ranging from $c_1 \approx 3.18$ to
$c_{\infty} = \pi^2/6$. The conductance is measured in units of
$2e^2/h$.  The second term in Eq.\ (\ref{eq:WL}) is the usual
weak-localization correction for non-interacting particles
\cite{BarangerMello}, the third term is the interaction correction.
Within the model (\ref{eq:model}) there is no dephasing, which would
give rise to a suppression of weak localization. The origin of the
absence of dephasing is that all interference occurs
inside the dot, and the interaction Hamiltonian does not produce
excitations during the course of such motion.  In the presence of
spin-orbit scattering (symplectic ensemble), the corrections to the
conductance [second and third term on the r.h.s.\ of
Eq.\ (\ref{eq:WL})] are multiplied by $-1/2$. Hence, in this case,
Coulomb interactions enhance the conductance.
We expect that the conductance distribution saturates when 
$\pi T$ becomes comparable to the level broadening $N \Delta/\pi$.
Hence for low temperatures, the interaction correction to $\langle
G \rangle$ remains small as $1/N$, and there is no
transition to the regime of real Coulomb blockade. 

(iii).
The capacitive interaction in the Hamiltonian (\ref{eq:model}) enhances
the mesoscopic conductance fluctuations. In the unitary ensemble, the conductance $G$ shows sample-to-sample fluctuations with variance
\begin{eqnarray}
  \mbox{var}\, G &=& {N \Delta \over 96 T} + {c_N \over N} {N^2 \Delta^2 \over 32 \pi^2 T^2}
  \ \ \mbox{for $T \gg N \Delta$},
  \label{eq:UCF}
\end{eqnarray}
with $c_N \approx 6.49$ for $N \gg 1$. In the absence of a
magnetic field, the fluctuations are larger by a factor $2$, while the
effect of strong spin-orbit scattering is to reduce $\mbox{var}\, G$ by
a factor $4$.  
The first term on the r.h.s.\ of Eq.\ (\ref{eq:UCF}), which represents the conductance fluctuations for non-interacting electrons, is derived in the
limit of large channel numbers, $N \gg 1$, and agrees with the result obtained by Efetov \cite{Efetov}.
The second term is the interaction correction. 

(iv).
The above results are for ideal ballistic point contacts and for
ergodic quantum dots.  The presence of a weakly scattering impurity in
the point contact with reflection amplitude $r$ gives rise to a
correction to the average conductance that behaves as
\begin{equation}
  \delta G = - {N^2 \over 2 \pi} c_N |r|^2 \left({2 N E_C \gamma \over \pi^2 T} \right)^{1/2 N} \sin {\pi \over 4 N},
\end{equation}
where $c_1 \approx 5.32$ and $c_N \to 4$ for $N \gg 1$. For $N=1$ this
result has been obtained previously in Ref.\ \onlinecite{FM}. However,
not only an impurity in the contact, but also any other direct
scattering process that acts on a time scale shorter than the Coulomb
time $t_c \propto 1/E_C$ (like scattering from short trajectories
connecting the point contacts), has the same effect on the
conductance.  The effect of all direct processes is summarized by the
replacement of the reflection probability $|r|^2$ by the difference
$|r|^2 - |t|^2$ of the probabilities for direct reflection and
transmission, respectively. In particular, we expect that as a result
of interactions, the conductance of an open quantum dot with a direct
path connecting source and drain contacts {\em increases} as the
temperature $T$ is lowered.

Let us now proceed with the formulation of the theoretical framework
used for the derivation of these results. The conductance
is computed from the current-current correlator in imaginary time,
\begin{mathletters} \label{eq:conductance}
\begin{eqnarray}
  G &=& {1 \over 2 T} \int_{-\infty}^{\infty} dt\,
        \Pi \left(it + {1 \over 2T} \right), \\
  \Pi(\tau) &=& \langle T_{\tau} I(\tau) I(0) \rangle.
\end{eqnarray}
\end{mathletters}%
We linearize the spectrum in the leads. In each lead there are $N$ channels,
labeled $j=1,\ldots,N$ for lead $1$ and $j=N+1,\ldots,2N$ for lead $2$.
The current $I$ is defined in the leads, just outside the quantum dot,
\begin{eqnarray}
  I &=& {e \over 4 \pi} \sum_{j=1}^{2N} \sum_{\sigma = \pm} \nu(j)
  \nonumber \\ && \mbox{} \times \left(
        \psi^{\dagger}_{Lj\sigma}(x) \psi^{\vphantom{\dagger}}_{Lj\sigma}(x)  -
        \psi^{\dagger}_{Rj\sigma}(x) \psi^{\vphantom{\dagger}}_{Rj\sigma}(x)
  \right),
\end{eqnarray}
where $\psi_{Lj\sigma}^{\dagger}$ ($\psi_{Rj\sigma}^{\dagger}$) and
$\psi_{Lj\sigma}$ ($\psi_{Rj\sigma}$) denote creation and annihilation
operators for left (right) moving particles with spin $\sigma$ in channel $j$,
and $\nu(j) = 1$ if $1 \le j \le N$ and $-1$ else. (We have chosen
units such that the Fermi velocity $v_F = 1/2 \pi$.) We take the limit
$x \uparrow 0$ at the end of the calculation.

The dynamics of the quantum dot is dealt with in the effective action
method of Ref.\ \onlinecite{AG}. Following Ref.\ \onlinecite{AG}, the
charge $Q$ of the quantum dot in the interaction Hamiltonian ${\cal
H}_C$ is replaced by the charge $Q_L$ of the leads. Such a replacement
is allowed, because the total number of particles (in the dot and in
the leads) is conserved. Next, the degrees of freedom of the quantum
dot are integrated out, in favor of a ``mirror'' copy of the leads at
$x > 0$, and an effective action ${\cal S}_{\rm eff}$, to be defined below. The
Hamiltonian of the system now reads 
\begin{mathletters} \label{eq:HamLin}
\begin{eqnarray}
  {\cal H} &=& {i \over 2 \pi} \sum_{j=1}^{2 N} 
               \sum_{\sigma = \pm} 
               \int_{-\infty}^{\infty} dx
  \left( \psi_{Lj\sigma}^{\dagger} \partial_x \psi_{Lj\sigma}^{\vphantom{\dagger}} - \psi_{Rj\sigma}^{\dagger} \partial_x \psi_{Rj\sigma}^{\vphantom{\dagger}} \right) 
  \nonumber \\ && \mbox{} + {E_{C} \over 2 e^2} (Q_L + ne)^2, \\
  Q_L &=& e \sum_{j=1}^{2 N} \sum_{\sigma=\pm} \int_{-\infty}^0 dx\, 
  : \psi_{Lj\sigma}^{\dagger} \psi_{Lj\sigma}^{\vphantom{\dagger}} + 
    \psi_{Rj}^{\dagger} \psi_{Rj\sigma}^{\vphantom{\dagger}} :.
\end{eqnarray}
\end{mathletters}%
The scattering of particles from the quantum dot is represented by the
imaginary-time effective action ${\cal S}_{\rm eff}$ acting at $x=0$,
\begin{eqnarray*}
{\cal S}_{\rm eff} &=& \sum_{i,j,\sigma,\sigma'} \int_0^{1/T} d\tau_1\, d\tau_2\, [\bar\psi_{Li\sigma}(\tau_1;0) + \bar\psi_{Ri\sigma}(\tau_1;0)]
  \nonumber \\ && \mbox{} \times
L_{ij;\sigma\sigma'}(\tau_1 - \tau_2)\, [\psi_{Lj\sigma'}(\tau_2;0) + \psi_{Rj\sigma'}(\tau_2;0)], 
\end{eqnarray*}
where $\psi(\tau;x) = e^{{\cal H} \tau} \psi(0;x) e^{- {\cal H} \tau}$
and $\bar \psi(\tau;x) = \psi^{\dagger}(-\tau;x)$. The kernel $L_{ij;\sigma\sigma'}(\tau)$ is a hermitian matrix, related to the scattering matrix $S_{ij;\sigma,\sigma'}$ of the quantum dot \cite{AG},
\begin{eqnarray}
 L(\omega_n) &=& \int_0^{1/T} e^{i \omega_n \tau} L(\tau) d\tau 
  \nonumber \\ &=& {1 \over 4 \pi i} {1 - S(\omega_n) \over 1 + S(\omega_n)} - {1 \over 4 \pi i}\, \mbox{sgn}(\omega_n), \label{eq:LS}
\end{eqnarray}
where $\omega_n = (2n+1) \pi T$ is the Matsubara frequency. Note that
in Matsubara representation, the kernel $L(\omega_n)$ satisfies
$L(\omega_n) = L(-\omega_n)^{\dagger}$. 
The current correlator $\Pi(\tau)$ is now calculated as a thermal
average with respect to the Hamiltonian ${\cal H}$ and with the
effective action ${\cal S}_{\rm eff}$,
\begin{eqnarray} \label{eq:average}
\Pi(\tau) &=& {\langle T_{\tau} I(\tau) I(0) e^{{\cal S}_{\rm eff}} \rangle_{\cal H} \over \langle e^{{\cal S}_{\rm eff}} \rangle_{\cal H}}, 
\end{eqnarray}

To compute the thermal average of the interacting system defined by
Eqs.\ (\ref{eq:conductance})--(\ref{eq:average}), we first bosonize the
one-dimensional Hamiltonian ${\cal H}$ of Eq.\ (\ref{eq:HamLin}). In
this way the interaction becomes quadratic in terms of the boson
fields, and can be dealt with exactly. To account for the scattering
from the quantum dot, which is represented by the effective action
${\cal S}_{\rm eff}$, we perform an expansion up to second order in
powers of the scattering matrix $S$. To be precise, we first expand
$\Pi(\tau)$ in powers of the action ${\cal S}_{\rm eff}$, and then we
expand ${\cal S}_{\rm eff}$ in powers of the scattering
matrix $S$ using that up to order $S^2$ we have $2 \pi i L(\omega_n) =
-S(\omega_n)$ if $\omega_n > 0$ and $S^{\dagger}(-\omega_n)$ otherwise,
see Eq.\ (\ref{eq:LS}).  Note that in Ref.\ \onlinecite{AG} the kernel
$L$ was used as the expansion parameter.  The expansion in the
scattering matrix $S$ instead of the kernel $L$ guarantees that in the
absence of interactions mesoscopic fluctuations are fully accounted
for. Only corrections to the conductance that depend on the interplay
of interactions and mesoscopic fluctuations are taken into account
perturbatively. The perturbation expansion is valid in the
regime $T \gg N \Delta$.  We find in this way
\bleq
\begin{eqnarray}
  G &=& {N \over 2} - {\pi T \over 8} \int_{0}^{\infty} d t_1\,
  d t_2 \sum_{k,l,\sigma,\sigma'}
  S^{*}_{lk;\sigma\sigma'}(t_1) 
  S^{\vphantom{remove this superscript}}_{lk;\sigma\sigma'}(t_2) 
  \nu(k) \nu(l)
  \left\{
  \vphantom{ \left( {\sinh \over \sinh } \right)^{1/2N}}
    {t_2 - t_1 \over \sinh[(t_2 - t_1) \pi T]}
  + T \sin {\pi \over 4 N}
  \right. \nonumber \\ && \left. \mbox{} \times
  \int_{t_c}^{\infty} d s
  { 2 s + t_2 + t_1 \over \sinh[(s + t_2) \pi T]
  \sinh[(s + t_1) \pi T]}
  \left( {
  \sinh[(t_2 + t_1 + s + t_c) \pi T]
  \sinh[(s - t_c) \pi T]
  \over
  \sinh[(  t_1 + t_c) \pi T]
  \sinh[(  t_2 + t_c) \pi T]
  } \right)^{1/4N}
  \right\} . \label{eq:result}
\end{eqnarray}
Here $t_c = \pi/2 N E_C \gamma$, $\gamma$ being the Euler constant. The
real-time or Lehmann representation of the scattering matrix is defined
through 
$
  S(\omega_n) = \int_0^{\infty} S(t) e^{- \omega_n t},\ \ \omega_n > 0.
$

Equation (\ref{eq:result}) contains the conductance of a specific
sample in the presence of the interaction (\ref{eq:HamC}), up to second
order in the scattering matrix $S(t)$ of the quantum dot. The first
term between brackets is nothing else than the Landauer formula for the
conductance of the non-interacting system. The remaining term is the
interaction correction. The perturbation theory in powers of the
scattering matrix is arranged in such a way, that we handle the
non-interacting ``pole'' contributions exactly, and do a perturbation
theory in the interacting ``branch-cut'' terms.
The sample-specific conductance (\ref{eq:result}) was obtained by a {\em
thermal} average of the interacting system. Next, we need to take
a {\em mesoscopic} average over an ensemble of quantum dots. Such an ensemble
can be obtained by varying e.g.\ the shape, impurity configuration, or the
Fermi energy of the quantum dot. The correlators of the scattering
matrix for such a mesoscopic ensemble have been studied extensively in
the literature \cite{BReview}. Although the two-point correlator of
scattering matrix elements is known exactly \cite{VWZ}, for our
purposes, within the regime $T \gg N \Delta$ where the expansion (\ref{eq:result}) is
valid, it is sufficient to use the asymptotic large-$N$ formulas for
these correlators. In the absence of spin-orbit scattering, the
scattering matrix is diagonal in spin space, $S_{ab;\sigma\sigma'} =
S_{ab} \delta_{\sigma\sigma'}$. For the unitary ensemble (presence of a
time-reversal symmetry breaking magnetic field), we have
\begin{mathletters} \label{eq:Scorr}
\begin{eqnarray}
  \langle S^{*}_{a b}(t) S^{\vphantom{remove this superscript}}_{a'b'}(t') \rangle &=& \delta_{a a'} \delta_{b b'}
  {\Delta \over 2 \pi} e^{-N \Delta t/\pi} \delta(t' - t),
\end{eqnarray}
while in the presence of time-reversal symmetry (orthogonal ensemble) we find
\begin{eqnarray}
  \langle S^{*}_{a b}(t) S^{\vphantom{remove this superscript}}_{a' b'}(t') \rangle &=& (\delta_{a a'} \delta_{b b'} +
  \delta_{a b'} \delta_{b a'}) 
  {\Delta \over 2 \pi} e^{-(2N+1) \Delta t/2 \pi} \delta(t' - t).
\end{eqnarray}
With spin orbit scattering in the presence of
time-reversal symmetry (symplectic ensemble), we have
\begin{eqnarray}
  \langle S^{*}_{a b;\sigma \tau}(t) S^{\vphantom{remove this superscript}}_{a' b';\sigma' \tau'}(t') \rangle &=& (\delta_{a a'} \delta_{b b'} \delta_{\sigma \sigma'} \delta_{\tau \tau'} - 
  \delta_{a b'} \delta_{b a'} (\sigma_y)_{\sigma \tau'} (\sigma_y)_{\tau \sigma'}) 
  {\Delta \over 2 \pi} e^{-(4N-1) \Delta t/4 \pi} \delta(t' - t),
\end{eqnarray}
\end{mathletters}%
where $\sigma_y$ is the Pauli matrix.  For interaction corrections to
leading order in $N \Delta/T$, higher order correlators of $S$ can be
taken Gaussian.  The ensemble average of Eq.\ (\ref{eq:result}) with
the help of the correlators (\ref{eq:Scorr}) yields the
weak-localization correction (\ref{eq:WL}) and the conductance
fluctuations (\ref{eq:UCF}).  (The observation that $\langle G \rangle
= N/2$ in the unitary ensemble follows from symmetry properties of the
ensemble of scattering matrices, and holds to arbitrary order in
perturbation theory.) The theory of Ref.\ \onlinecite{FM}, where the
effect of weakly reflecting impurities in the contacts was
considered, is recovered by setting $S_{ij;\sigma\sigma'}(t) = r_j
\delta(t) \delta_{ij} \delta_{\sigma \sigma'}$, $r_j$ being the
reflection amplitude in channel $j$.

For spinless particles (which might e.g.\ be realized in spin-polarized
quantum dots), Eq.\ (\ref{eq:result}) remains valid, provided one
substitutes $N \to N/2$. However, if the leads contain only one channel
($N=1$), an extra $n$-dependent term $\delta G$, has to be added,
\begin{eqnarray}
  \delta G &=&
  {\pi T^2 \over 4} 
  \int_{t_c}^{\infty} d s
  \int_{0}^{\infty} dt_1 dt_2 
  {\mbox{Re}\, e^{-2 \pi i {n}} (2 s + t_1 + t_2) \left[S_{12}(t_1) S_{21}(t_2) - S_{11}(t_1) S_{22}(t_2) \right]\over \sqrt{ 
  \sinh[(t_c+t_1) \pi T] \sinh[(t_c+t_2) \pi T]
  \sinh[(t_c+s+t_2+t_1) \pi T] \sinh[(s - t_c) \pi T]}}.
  \label{eq:single}
\end{eqnarray}
\eleq\noindent
In this case the weak localization correction to the conductance is enhanced,
\begin{equation}
  \langle G \rangle = 
  {1 \over 6} + {\Delta \over 16 \pi T} \ln {t_c \pi T \over 2 c},
\end{equation}
where $c \approx 0.36$.
Eq.\ (\ref{eq:single}) bears explicit reference to the particle number $n$,
and induces an explicit $n$-dependence in the conductance fluctuations,
\begin{eqnarray}
  \langle G(n) G(n') \rangle &=& {c \Delta^2 \over 128\, \pi^2T^2} \cos[2 \pi (n-n')]
 \ln^2(t_c \pi T) \nonumber \\ && \mbox{} + \mbox{$n$-independent terms},
\end{eqnarray}
where $c =5/4$ ($1/2$) in the presence (absence) of time-reversal
symmetry.  In our perturbation theory, such an explicit $n$-dependence,
which reflects the discreteness of charge, is absent for particles
with spin. The origin for the appearance of this extra term for
spinless particles and $N=1$ is the same as in Ref.\ \onlinecite{FM}.
For spin $1/2$ electrons, effects of the discreteness of charge appear
in general in $4N$th order in perturbation theory in $S$.

We close with a remark on the issue of dephasing, which plays an
important role in the experiments on open quantum dots\cite{Marcus}.
While the experiments indicate that the dephasing time $\tau_{\phi}$
diverges as $T^{-1}$ as the temperature $T \to 0$, the theoretical
prediction for a closed quantum dot is $\tau_{\phi} \propto T^{-2}$
\cite{SAI}. As our model (\ref{eq:model}), which takes the effect of
open contacts into account, does not give rise to significant dephasing
effects, we believe that the fact that a quantum dot is open by itself
does not lead to additional dephasing, and cannot explain the puzzling
discrepancy between experiment and theory.

It is a pleasure to acknowledge discussions with B.\ L.\ Altshuler,
C.\ W.\ J.\ Beenakker, B.\ I.\ Halperin,
C.\ M.\ Marcus, and Y.\ Oreg. PWB is supported by the NSF under
grants no.\ DMR 94-16910, DMR 96-30064, and DMR 94-17047 and IA is
A.\ P.\ Sloan research fellow.

\ecols
\end{document}